\documentclass[prb,preprint]{revtex4-1}

\usepackage{amsmath}  % needed for \tfrac, \bmatrix, etc.
\usepackage{amsfonts} % needed for bold Greek, Fraktur, and blackboard bold
\usepackage{graphicx} % needed for figures
\usepackage{float}    % allow exact placement of figures

\begin{document}
%\linenumbers
%\modulolinenumbers[5]

\title{A simple demonstration of shear-flow instability.}

\author{Tom Howard}
\email{tom.howard@metoffice.gov.uk} % optional
\affiliation{Met Office, FitzRoy Road, Exeter, EX1 3PB, United Kingdom.}
% Please provide a full mailing address here.
\author{Ana Barbosa Aguiar}
\affiliation{Met Office, FitzRoy Road, Exeter, EX1 3PB, United Kingdom.}
% See the REVTeX documentation for more examples of author and affiliation lists.

\date{\today}

\begin{abstract}
We describe a simple classroom demonstration of a fluid-dynamic instability. The demonstration requires only a bucket of water, a piece of string and some used tealeaves or coffee grounds.
We argue that the mechanism for the instability, at least in its later stages, is two-dimensional barotropic (shear-flow) instability and we present evidence in support of this.  We show results of an equivalent
basic two-dimensional numerical non-linear model, which simulates behavior comparable to that observed in the bucket demonstration. Modified simulations show that the instability does not depend
% solely 
on the curvature of the domain, but rather on the velocity profile.
\end{abstract}
% \keywords{instability, barotropic, laboratory, classroom demonstration}

\maketitle

\section*{Videos}
The manuscript cites  video clips which are available password-protected on the vimeo video-sharing site.
We strongly encourage readers to watch these video clips, in particular the following two, as they add a lot to the description given by the text and figures.

The experiment:
VideoS1  available at \href{https://vimeo.com/278481176}{\texttt{https://vimeo.com/278481176}} (password: Welcome123)  % physical demo

The numerical simulation (slow motion):
VideoS2 available at 
\href{https://vimeo.com/399593365}{\texttt{https://vimeo.com/399593365}} (password: Welcome123),   %  4-panel anim (.../28/...)

This article may be downloaded for personal use only. Any other use requires prior permission of the author and AIP Publishing. This article appeared in American Journal of Physics 88, Issue 12, 1041 (2020) and may be found at \href{https://doi.org/10.1119/10.0002438}{\texttt{https://doi.org/10.1119/10.0002438}}

\section{Introduction}
%\introduction  %% \introduction[modified heading if necessary]

Shear-flow instability is a fundamental process in fluid dynamics, and is associated with the destruction
of parallel laminar flow.
As such, it can be seen as
fundamental to the onset and existence of both two- and three-dimensional turbulence.\citep[e.g.][]{phillips1969shear}
Geophysicists usually use the term ``barotropic instability'' in the context of the instability of a horizontally-sheared fluid on a rotating planet.
The word ``barotropic'' distinguishes this from baroclinic instability, in which horizontal temperature gradients play a key role.
Baroclinic instability is the major large-scale weather-generating process of
the mid-latitudes (the region between the tropics and the polar circles), whereas
barotropic instability is important as an instability
mechanism for jets and vortices.
Both instabilities may be associated with the polygonal shapes often observed in the atmospheres of rotating planets.\citep[e.g.][]{Kossin, Adriani}
% Furthermore, the study of barotropic instability may lead to useful insights into the baroclinic problem \citep{Vallis}.

% In an intriguing recent article, Lee et al.\cite{lee2018sinking} have suggested that shear-flow instability may be responsible for the formation of the waves which can 
% be observed in the sinking bubbles seen in a settling pint of stout beer.

The instructional value of laboratory demonstrations is well-recognized, both
in the context of geophysical fluid dynamics in particular,\cite{Illari, Marshall, mackin2012effectiveness} and
in the context of fluid dynamics in general (Shakerin\cite{shakerin2018fluids} and references therein).
% (see Ref.~\onlinecite{shakerin2018fluids} and references therein).
%   Shakerin\cite{shakerin2018fluids} gives instructions for four very low-budget fluids demonstrations including one which illustrates two types of flow instability
%   (Rayleigh-Taylor instability and Saffman-Taylor instability).
%
A classroom
demonstration (e.g. \cite{BIVideo})  of baroclinic instability can be prepared based on the
``dishpan'' experiments of the mid-twentieth century.\citep[e.g.][]{Fultz}
To study barotropic instability in the laboratory, sophisticated experiments somewhat analogous to the
dishpan experiments can be constructed using differentially rotating boundaries.\citep[e.g.][]{Aguiar, Konijnenberg, Nezlin}
The shear is
introduced due to the differential rotation of concentric sections of the bottom of the container of the working fluid.
% and communicated to the fluid through the bottom boundary layer. 
However, such an apparatus is beyond the budget of most classroom teaching of geophysical fluid dynamics.

Reynolds~\cite{Reynolds} describes experiments with a horizontal glass tube containing carbon disulphide
overlain by water. Carbon disulphide is a highly toxic liquid with a density approximately 25\%  higher
than water at room temperature. By tilting the tube, Reynolds induced a counterflow in the two liquids
and this led to instability at the shear interface when the velocity difference was large enough. 
%Reynolds
%described this as a ``very pretty experiment'' which ``completely answered its purpose''. 
A version of this
apparatus, using a carefully-established saline/fresh water step to create the density difference,
% (and thus more suited to twenty-first century safety-consciousness), 
is a well-loved teaching aid 
% in the G.K.Batchelor Laboratory in the Department of Applied Mathematics and Theoretical Physics 
at the
University of Cambridge.\cite{KHVideo} However, the equipment
is quite specialized and the set-up procedure is painstaking and time-consuming.
One simpler classroom option is to introduce shear by moving a solid boundary through a fluid --- for example
a cylinder with vertical axis, partially submerged in water, may be towed sideways.
Food dye applied to the cylinder before immersion, or
mica powder
mixed into the water, exposes the vortices which appear in the wake.
    Kelley and Ouellette\cite{kelley} describe a laboratory approach aimed at undergraduate students, who use an
    electromagnetic technique to drive Kolmogorov flow (a type of cyclic shear flow which exhibits shear-flow instability under some forcing parameters) in a thin fluid layer, and measure it
    quantitatively with a webcam. The authors estimate costs of the order of \$~500.
    Vorobieff and Ecke\cite{vorobieff1999fluid} present an apparatus based on a tilted gravity-driven soap tunnel, with which a variety of fluid-dynamic phenomena,
    including shear instability, can be demonstrated. Their estimated costs were of the order of \$1000 (in 1999).

Here we describe an instance of an instability that can be easily reproduced in a
bucket and used as a very low-budget demonstration, which we have found to engage a wide range of student groups. We argue that the instability, at least in its later stages,
is consistent with the two-dimensional shear flow instability exhibited by a basic numerical model.

\section{Brief review of the mechanism of shear-flow instability}
\label{mechanism}

To introduce the mechanism we begin with a two-dimensional conceptual model consisting 
of two regions of incompressible fluid of constant uniform density (let's call them the North and South parts)
each in uniform flow in opposite directions like two counterflowing carriageways of traffic. Nothing divides them; their 
interface is initially a straight vertical plane.
For simplicity in this section we do not consider a curved shear layer as seen in the bucket. We show in subsection~\ref{straightChannel} that 
the curvature is not an essential feature of our demonstration. 

\vspace{5mm}
%The key feature of shear flow instability is this: \emph{interaction of the two parts can exchange momentum between them and release kinetic energy, which is
The fundamental cause of shear flow instability is this: \emph{interaction of the two parts can exchange momentum between them and release kinetic energy, which is
then available to drive a more complex (sometimes chaotic) behavior in the region of the interface.}
\vspace{5mm}

\begin{figure}[H]
% \noindent\includegraphics[width=40pc,angle=0]{fig_HW.eps}\\
  \noindent\includegraphics[width=40pc,angle=0]{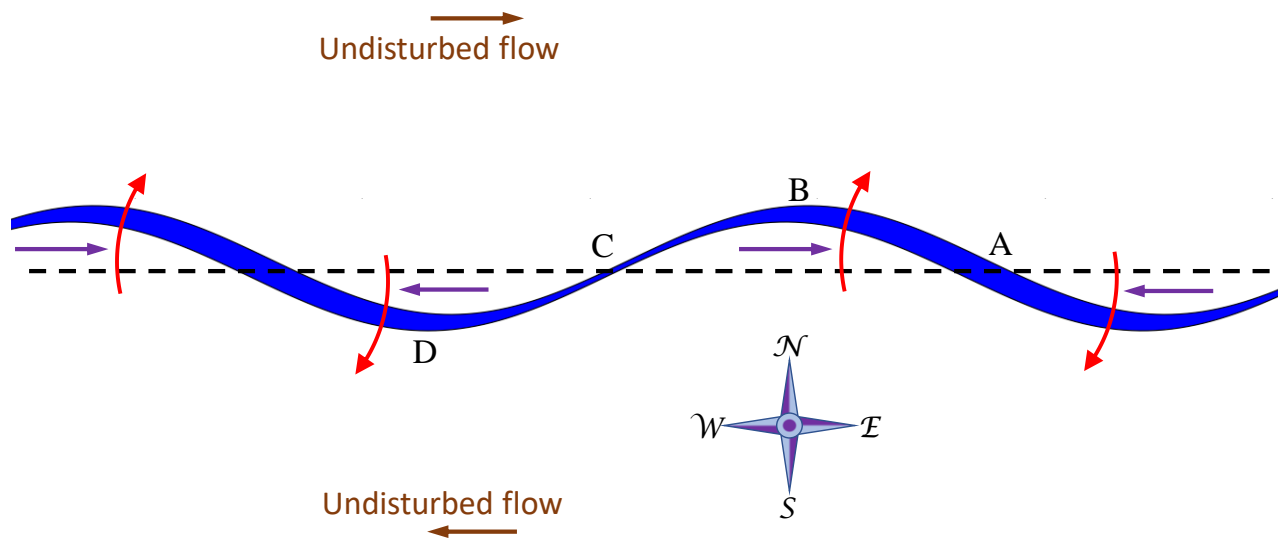}\\
%x = np.linspace(0,14, 200)
%plt.clf()
%plt.fill_between(x, np.cos(x), np.cos(x+0.2)+0.3)
%plt.ylim((-1.5, 1.5))
  \caption{
Schematic diagram showing the positive feedback of a small sinusoidal disturbance to the interface of two counterflowing regions of fluid.
The compass rose is for ease of reference in the main text; other orientations are, of course, possible.
}
  \label{figBatch}
\end{figure}

The details of the mechanism are most easily understood in terms of a derived property of fluid motion called vorticity. 
Mathematically, vorticity is the curl of the velocity field.
In a two-dimensional flow context, we can regard the vorticity as a scalar, because its direction is always the same (normal to the plane of the motion).
Shapiro\cite{shapirovideo} offers a striking physical interpretation of vorticity, which we paraphrase here:
``The vorticity is a measure of the moment of momentum of a small fluid particle about its center of mass.
Suppose that you had some very complicated [two-dimensional] motion in a liquid  and that it were possible by magic  suddenly
to freeze a small sphere of that liquid into a solid.
During the freezing the moment of momentum would be conserved.
The vorticity of the fluid before freezing is exactly twice the angular velocity of the solid sphere just after its birth.''
This vivid interpretation may sound fanciful, but the mental image it conjures is not so different from what we see 
% (albeit on a large scale) 
in the development of shear-flow instability: 
vorticity (which is initially present as a continuous shear at the interface) is rapidly converted 
by the instability
into the rotation of a number of vortices along the line of the interface.
The mechanism is illustrated in Fig.~(\ref{figBatch}), which is adapted from Batchelor.\cite{batchelor1967}
The process begins with undisturbed counterflow and a straight interface as described above, shown by the dashed straight line (this line would have a monotonic curve in the case of the bucket demonstration).
We introduce a small sinusoidal disturbance as shown. 
It can be shown that disturbed vorticity south of the dashed line (for example at D) 
% implies (``induces'') 
induces
eastward movement in the vorticity north of the dashed line, 
and likewise vorticity north of the dashed line (for example at B) 
% implies (``induces'') 
induces
westward movement in the vorticity south of the dashed line.
Both of these effects concentrate vorticity at points like A, giving a positive feedback, with energy provided by the kinetic energy of the counterflowing parts.
The arrows indicate
the direction of the movement of the vorticity induced by the disturbance,
and show both the accumulation of
vorticity at points like A (purple arrows) and the general rotation about points like A (red arrows) which together lead to growth of the disturbance.
The accumulation of vorticity (at points like A) and depletion of vorticity (at points like C) is also
indicated by the thickness of the wavy line. A comprehensive discussion of this case is given by Batchelor.\cite{batchelor1967}
% The varying thickness of the wavy line also indicates the accumulation of vorticity (at points like A) and dissipation of vorticity (at points like C)
% The accumulation (dissipation) of vorticity at points like A (C) is also indicated by the thickening (thinning) of the wavy line.

In this simplified configuration
the sinusoidal disturbance grows 
%around nodes  (like A and C for example) with fixed spatial locations. Likewise,
% ultimately, distinct vortices form around fixed spatial locations (like point A for example).
around nodes  (like point A for example) with 
fixed spatial locations where distinct vortices will, ultimately, form. 
Extending the traffic analogy, a typical vortex centre is like a fixed point on the road, with traffic sweeping past one way on one side, and the other way on the other side.

In a more general case where the velocities are not equal and opposite, for example in the demonstration which we describe next,  the vortices move along the line of an interface at a speed intermediate 
between the faster-moving fluid on one side and the slower-moving fluid on the other side of the interface.

\section{Demonstration}\label{Demonstration}
\subsection{Description}\label{DemonstrationDescription}
\begin{figure}[t]
%  \noindent\includegraphics[width=19pc,angle=0]{fig-GS.png}\\
  \noindent\includegraphics[width=19pc,angle=0]{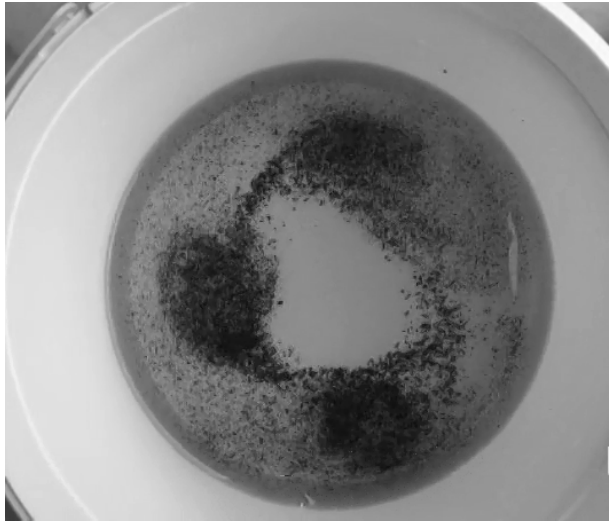}\\
  \caption{Snapshot of three vortices in a bucket, visualized using tealeaves.}
  \label{FigBucketSnapshot}
\end{figure}
A bucket (ours has an internal diameter of 180 mm) is filled with tap water to a depth of about 30 mm.
A few used tealeaves or coffee grounds are added as a simple flow tracer. % visualization medium.
A piece of string is attached to the center of the bucket handle and the bucket lifted a little 
so that it is suspended on the string.
A vigorous angular impulse (a flick) is applied to the handle using the fingers, and the bucket
allowed to spin for a few turns. This sets the water at and near the wall in rotation.
The bucket is then set back down, arresting its rotation (and attenuating the motion of the water nearest to the wall).
 Typically,
within a few seconds, the axisymmetry of the resulting flow breaks and a number of
smaller vortices (typically two, three or four) appear --- see Fig.~(\ref{FigBucketSnapshot}) 
and our first video clip VideoS1,
available at 
\href{https://vimeo.com/278481176}{\texttt{https://vimeo.com/278481176}} (password: Welcome123).  % physical demo
These smaller vortices, which 
rotate in the same sense as the initial rotation of the bucket, will typically merge into
a near-axisymmetric single vortex in a few more seconds, before the motion dies completely. 
With careful observation, the
vortices can also be seen on the top surface, for example by sprinkling buoyant glitter onto the water. 
However the tealeaves 
(which have a small negative buoyancy and thus illustrate the flow near the bottom boundary) 
have the well-recognized tendency to converge in the vortices,\citep{tandon2010einstein}
and this provides an impactful visualization.
Incidentally, we came upon our demonstration by accident whilst playing with a version of the well-known ``Einstein's Tea Leaves'' 
demonstration,\citep{tealeavesfilm}      which is widely used in teaching the behavior of air near the surface of atmospheric pressure systems.\citep{tandon2010einstein}

% long version: attempt to correct a misunderstanding by one of the BAMS reviewers ( /home/h03/hadto/bucket/paper/BAMS/submission01/rev3_bams-xi18.pdf ):
% The three vortices illustrated in Fig.~(\ref{FigBucketSnapshot}) all rotate in the same sense. 
% This is distinct from the tripolar vortex structure
% studied by, for example, van~Heijst\cite{vanHeijst} in which the three vortices are all aligned, with the center or core vortex
% rotating in the opposite sense to the two satellite vortices.
% % also studied by ./polvani+carton-GAFD-1990.pdf

\begin{figure*}[t]
% \noindent\includegraphics[width=25pc,angle=0]{fig-FX.png}\\
% \noindent\includegraphics[width=25pc,angle=0]{ig-FX.png}\\
  \noindent\includegraphics[width=27pc,angle=0]{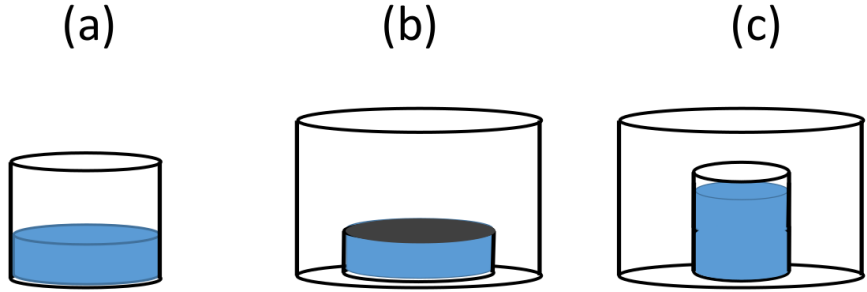}\\
  \caption{ Schematic diagram showing cross-sectional views of alternative experimental setups which all
exhibit similar instability. (a) Initial setup. (b) As (a) but in a larger outer bucket facilitating the addition
of a rigid lid (see main text). (c) Taller and narrower. In this setup the large outer bucket helps to reduce
wobble. Bucket handles and string not shown.}
\label{figConfigs}
\end{figure*}

A snapshot of the demonstration is shown in Fig.~(\ref{FigBucketSnapshot}). The
initial flick was of the order of four revolutions per second and the rotation of the bucket was then
arrested by replacing it on the table after about two seconds. However, the behavior seems to be quite
robust to variations as long as the initial flick is strong enough. We have observed the instability in a taller, narrower configuration (120 mm
diameter by 100 mm deep) and in a setup with a rigid 
lid, arranged by filling a transparent cylindrical
plastic container (a cake box) with water, placing a larger bucket upside down over the box so that the
base of the bucket forms a lid to the box, and then inverting the whole. Atmospheric pressure
stops the water from draining out of the box (rather like a very squat manometer). Similar instability
behavior is seen in each case.
% These alternative configurations are shown schematically in Fig.~(\ref{figConfigs}), and links to video clips of 
% their behavior are given in Appendix~\ref{Links}.
These alternative configurations are shown schematically in Fig.~(\ref{figConfigs}), and video clips of 
their behavior are linked from section~\ref{Links}.
% The instability is not usually seen in a typical tea-leaves-in-a-cup demonstration. This is because in such a demonstration

\subsection{Curricular Context}

In a teaching context, this demonstration could, for example, be included in
a session introducing the general behavior 
of the mid-latitude atmosphere via the concept of fluid-dynamic instability.
The initial velocity profile (which is shown in Fig.~(\ref{figProfiles})) can be interpreted as a jet, whose instability leads to 
the production of several vortices. 
% Whilst this is clearly a very loose and highly incomplete model
  Whilst this is         a      loose and  incomplete model
of the mid-latitudes, it does provide an engaging, hands-on approach. Given the 
low cost, it is easy to provide a group of students with several buckets so that they can 
experiment and observe the behavior in small groups.
%In the early 20th century barotropic and baroclinic instability were rival hypotheses 
%for the explanation of the observed mid-latitude eddies \citep{James}, and so  
%the discussion then leads quite naturally to baroclinic instability.
%Having thus introduced the general concept of fluid-dynamic instability, 
% Later, baroclinic instability was recognized as the primary explanation, so 
%the session could then
%move on to a  demonstration of baroclinic instability such as the one described 
%by \citet{Marshall}.
In the early 20th century barotropic and baroclinic instability were rival hypotheses 
for the explanation of the observed mid-latitude eddies,\citep{James} and so  
having introduced the general concept of fluid-dynamic instability via the buckets,
the session can then
move on to a  demonstration of baroclinic instability such as the one described 
%by Ref.~\onlinecite{Marshall}.
by Marshall and Plumb.\cite{Marshall}
Alternatively, the activity could play a role in demystifying observations 
of polygonal patterns in the atmosphere of rotating planets.\citep[e.g.][]{Kossin, Adriani}

\vspace{5mm}
\subsection{Nature of the instability}\label{nature}
When the bucket is
arrested, the water in contact with the bucket is also arrested (in accordance with the no-slip condition).
So now the depth-averaged azimuthal velocity reaches a peak \emph{near} the wall but falls to zero \emph{at} the wall.

As discussed by Rabaud and Couder,\cite{rabaud1983shear} Coles\cite{coles1965transition}
observed patterns of ``rollers'' with their axes parallel to the rotation axis (Coles' Fig. 22(o)) in a Couette geometry (concentric independently-rotating cylinders), associated with
sudden starts and stops of the rotation of the outer cylinder, which created an inflection in the velocity profile. This description is strikingly similar to our experiment and observations.
% mentions:
% However, patterns of rollers with their axes parallel to the rotation axis were
% observed in the Couette geometry by Coles (1965). They corresponded to transitory
% states when sudden starts and stops of the rotation of the outer cylinder created an
% inflection in the radial profile of the velocity.

% @article{rabaud1983shear,
%   title={A shear-flow instability in a circular geometry},
%   author={Rabaud, M and Couder, Y},
%   journal={Journal of Fluid Mechanics},
%   volume={136},
%   pages={291--319},
%   year={1983},
%   publisher={Cambridge University Press}
% }
% 
% @article{coles1965transition,

\begin{figure}[t]
% \noindent\includegraphics[width=19pc,angle=0]{fig-FY-wider-e.png}\\
% \noindent\includegraphics[width=19pc,angle=0]{HowardFig03.eps}\\
% \noindent\includegraphics[width=27pc,angle=0]{fig-FY-improved.eps}\\
  \noindent\includegraphics[width=27pc,angle=0]{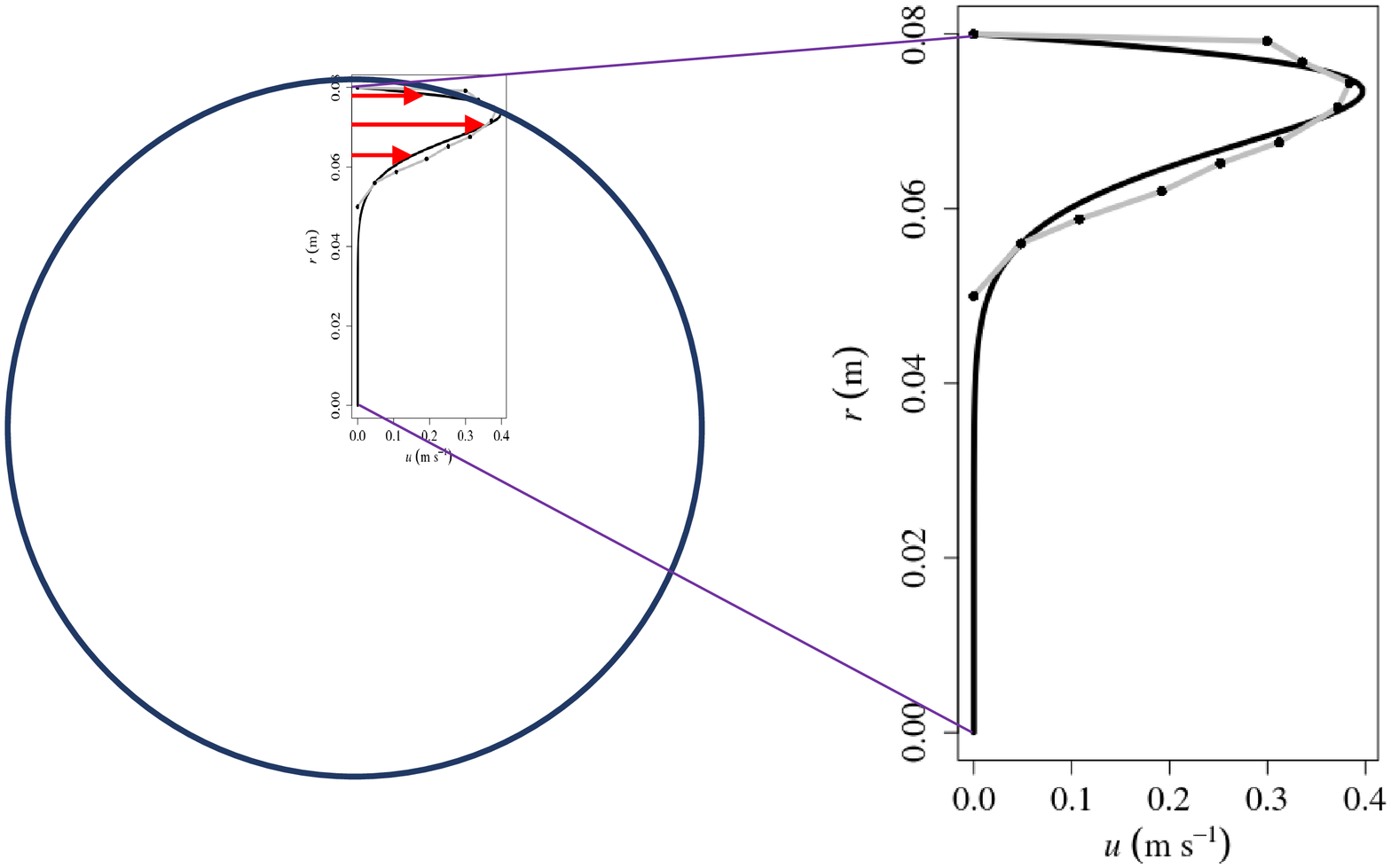}\\       
  \caption{ Estimated azimuthal velocity (\protect{$u$}) profile against radius (\protect{$r$}) from a typical experiment, before any instability was observable by eye,
is shown by the black dots connected by a gray line. Speeds were estimated by manually tracking
particles through several frames of a stop-motion film. Uncertainties in this profile are difficult to
estimate but an error of the order of 0.02~\protect{m~s$^{-1}$} is plausible. It was apparent, however, that all
velocities were positive at this stage. The point representing zero velocity at the outside wall (\protect{$r=0.08$~m}) is assumed, based
on the no-slip condition, rather than estimated from observations. The analytical profile used in the modelling
experiments (see section~\protect{\ref{sectModel}}) is shown by the continuous black line.
The large circle on the left-hand side of the plot represents the bucket seen from above, to clarify the context of the profile.
}
  \label{figProfiles}
\end{figure}

Our  crudely-measured velocity profile, from before any visible onset of instability, is shown in Fig.~(\ref{figProfiles}). 
The velocity profile at this stage satisfies the Rayleigh-Kuo criterion,\citep{andrews2010introduction} which is a necessary (though not sufficient)
condition for shear-flow instability.
% short version:
In our context this criterion is that the velocity profile must include an inflection point, i.e. a point at which the curvature
of the velocity profile changes sign.
% long version:
% This criterion is that the radial gradient of the vorticity must change
% sign at least once \citep[see for example][]{DrazinReid} or, equivalently, that the
% vorticity profile must include an extremum within the domain.
% The instability is not usually seen in a typical tea-leaves-in-a-cup demonstration. 
% This is probably because in such a demonstration the radial gradient of the vorticity does not change sign: to produce
% the instability a flick-start-stop initial rotation as described in subsection~\ref{DemonstrationDescription} is required.
    % , or, again, that the 
    % velocity must include an inflection point, i.e. a point at which the curvature
    % of the velocity profile changes sign at least once (see Drazin & Reid 1981). --- Only exact for the straight channel. For bucket need to include curvature term in the vorticicty calc
    % Or, satisfies the Rayleigh inflection point criterion 
    % These alternative forms of the criterion can be viewed as
    % q=-du/dy = vorticity  !! except this is not quite correct for the curved channel: need curvature term  q0 = -du/dr - u/r
    % dq/dy must vanish somewhere in the domian
    % Or ,equivalently
    % the velocity must include an inflection point: a point at which the curvature of u changes

In initial discussion with correspondents (see acknowledgments), various candidate processes were
suggested for the observed instability, including centrifugal instability generating turbulence 
% (as discussed above), % only in the long version!
followed by an up-scale energy transfer; an instability of the Ekman layer; and shear instability. 
%A necessary criterion for symmetric overturning instability is that the angular momentum 
%decreases outward, and this is the case for the profile shown, outside of the velocity maximum which occurs at $r \approx 0.0735$~m. 
Cullen (pers. comm.) noted that the observed behavior appears to be quasi-two-dimensional, and that such
flow would not be observed in the first place if it was unstable to three-dimensional disturbances. 
In the following section we show that a two-dimensional model (which admits no variation in the vertical direction)
exhibits behavior similar to that observed when the model is initialized with a profile based on the measured profile of Fig.~(\ref{figProfiles}).

\section{Numerical model}\label{NonLinearModel}  \label{sectModel}  %% Appendix A   This is the non-linear model description
\subsection{Description}\label{NonLinearModelDescription}
% A. Description B. Numerical modelling results and discussion C Modified#
We describe a non-linear two-dimensional numerical model of the instability seen in the bucket.
%The stream function and vorticity $\psi$ and $\zeta$ (defined in section~\ref{mechanism}) are scalars. Their values do not depend on the coordinate system in which they are evaluated. 
%They provide a convenient tool for a two-dimensional numerical model
The stream function, $\psi$, and vorticity, $\zeta$, are convenient tools for such a model.
For our purpose, each is a scalar function of space and time.
% Mathematically, the stream function, the 2-D Laplacian of which is the vorticity.
% Mathematically, the vorticity is the     2-D Laplacian of the stream function.
Mathematically, we define the stream function as a field, the 2-D Laplacian of which is the vorticity field.
Physically, this gives the useful properties that the instantaneous contours of the stream function (which may be familiar to some readers as ``streamlines'') 
are everywhere tangent to the instantaneous velocity vector, and that
regions of rapid movement can be identified by closely-packed streamlines.
The streamlines indicate the paths that particles in the flow would take if the flow did not evolve over time.
In our case the flow does evolve over time, so the stream lines are not identical to particle paths (see VideoS3).

The values of stream function and vorticity do not depend on the coordinate system in which they are evaluated.
 For our purpose it is convenient to use a polar coordinate system:
% Our non-linear model
% % Page "1 Feb 2018 Summary" in my notes
% is a two-dimensional 
we use a polar grid with
256 azimuthal grid points and 112 radial grid points. The integration procedure is:

\begin{itemize}
\item Given a vorticity field
\item Invert the Laplacian to obtain the stream function, $\psi$, given in polar coordinates by $u = - {\partial \psi}/{\partial r}$,
$v = {\partial \psi}/{r\partial \theta}$, where    $u$ is azimuthal velocity component, $v$ is meridional (i.e. radial) velocity component, 
$r$ is radius and $\theta$ is the azimuthal angle (in radians, defined positive in the direction of increasing $u$)
% and $\zeta = \nabla^2 \psi$.
\item Evaluate the components of velocity $(u, v)$ from the gradient of the stream function
\item Advect the vorticity field for one time-step. ``Advection'' refers to 
movement by the flow of the fluid; thus in this context ``advect'' the vorticity
means move the vorticity, using the velocity that we evaluated. We use a simple upwind explicit scheme with a small time-step to accomplish this.
\item Repeat
\end{itemize}

Inversion of the Laplacian is a linear problem that can be addressed by a Fourier decomposition in the
azimuthal direction. Owing to the linearity, the modes can be treated independently. This requires one
tri-diagonal matrix inversion for each mode.

We initialized our model with the smooth velocity profile shown by the black line in Fig.~(\ref{figProfiles}).
This velocity is a heuristic analytical function of the radius,  as follows:  $u(r)=u_0 \alpha \exp(-\alpha^{\gamma})$, where
$\alpha=(R-r)/w$, 
% $u$ is the azimuthal component of velocity, $r$ is the radius, 
$R$ is the outer radius, and $u_0, w, \gamma$ are tunable
parameters, $w$ being a scale width for the velocity jet. The fit was chosen by eye such that the peak azimuthal speed and the radius at which it occurred
matched well with the measured profile. We used $w=0.008$~m,  $u_0=1.05$~{m~s$^{-1}$}, $\gamma = 1.3$.
This gives a profile which is simple to work with, whilst maintaining consistency with the measurements to within their uncertainties.
A more complicated alternative would be to use piecewise linear interpolation (as shown by the gray lines) between each of the measurements, but 
this seems unjustified given the uncertainties.
The radial gradient of vorticity changes sign at about $r=0.0679$~meters for our fitted profile, and somewhere between $r=0.055$~m and $r=0.072$~m for the measured profile. 

We add a field
of small random vorticity `noise' to initialize any instability. Typical magnitudes (root-mean-square) of
the noise are set to be of the order of 0.1\% of typical values of the initial vorticity profile.
To sidestep the difficulty with the singularity at the pole we truncate the grid at a small but non-zero
radius and approximate the behavior of the fluid within this radius as a blob \label{blob} of material which is allowed to revolve in rigid
rotation around the pole but is otherwise identical to the rest of the fluid.
(We suggest that this is only a minor constraint on the motion, and we note that a version of the laboratory demonstration
not shown here including a comparable fixed rigid central cylinder of $\sim 20$~mm diameter exhibits very similar instability behavior.)
This leads naturally to a
Neumann boundary condition on $\psi$ for the $k=0$ mode (which represents the average) of the stream function
at the inner boundary:
\begin{equation}
\frac{\partial\psi_k}{\partial r} = -\frac{1}{2}\zeta_k r \ \ , \ \ \text{inner boundary}, \ \ (k = 0),
\label{eqNeumann}
\end{equation}
since
$\frac{1}{2}\zeta_k r \ \ (k = 0)$ is the azimuthal velocity of the rigid circular blob. Konijnenberg et al.~\cite{Konijnenberg} use a similar boundary condition (their equation 5.8).
For the other modes, we apply a Dirichlet boundary condition
at the inner boundary:
\begin{equation}
\psi_k = 0, \ \ \ \ \text{inner boundary}, \ \ (k \neq 0),
\end{equation}
which represents the constraint that no fluid can pass through the circular boundary of the rigid blob. At
the outer boundary we apply a Dirichlet boundary condition to all modes (including $k = 0$):
\begin{equation}
\psi_k = 0, \ \ \ \ \text{outer boundary, all }k.
\end{equation}
Again, for the $k \neq 0$ mode, this represents the constraint that no fluid can pass through the outer
circular boundary of the domain (the wall of the bucket), and in the case of the $k = 0$ mode, this effectively
sets the arbitrary zero for the stream function so that the solution is unique.

The inner boundary conditions for the vorticity are
% as two lines:
% \begin{displaymath}
% \frac{\partial \zeta_k}{\partial r} = 0 \ \ \ \ (k=0), \text{and}
% \end{displaymath}
% \begin{displaymath}
% \zeta_k = 0 \ \ \ \ (k \neq 0).
% \end{displaymath}

% as one line:
\begin{displaymath}
\frac{\partial \zeta_k}{\partial r} = 0 \ \ \ \ (k=0), \ \ \ \  \text{and} \ \ \ \ \zeta_k = 0 \ \ \ \ (k \neq 0).
\end{displaymath}

Treating the circular blob as rigid implies that its vorticity has a single value at any time. Our
implementation of Eq.~(\ref{eqNeumann}) ensures that this vorticity is the average of the vorticity of the fluid just
outside the blob (since the $k = 0$ mode represents the average).

We do not include any explicit model of diffusion. Our first-order upwind scheme is known to be diffusive, and one might
optimistically argue that this property of the two-dimensional numerical model 
mimics the damping influence of the bottom boundary in the bucket. However we have not made any 
quantitative comparison of these damping factors.
% ; we merely note that the model remains stable in the numerical sense.

% \subsection{}     %% Appendix A1, A2, etc.

% To facilitate sub-sampling of the estimated velocity profile shown by the dots in Fig.~(\ref{figProfiles}), we fitted by eye
% a heuristic analytical function to the profile, as follows:  $u(r)=u_0 \alpha \exp(-\alpha^{\gamma})$, where 
% $\alpha=(R-r)/w$, 
% $u$ is the azimuthal component of velocity, $r$ is the radius, $R$ is the outer radius, and $u_0, w, \gamma$ are tunable
% parameters, $w$ being a scale width for the velocity jet. The fit was chosen by eye such that the peak azimuthal speed and the radius at which it occurred
% matched well with the measured profile. We used $w=0.008$~m,  $u_0=1.05$~{m~s$^{-1}$}, $\gamma = 1.3$. 
% This fitted analytical profile is shown by the black line in Fig.~(\ref{figProfiles}).
% The radial gradient of vorticity changes sign at about $r=0.0679$~meters for this fitted profile, and somewhere between $r=0.55$~m and $r=0.072$~m for the measured profile. 
% see ~/bucket/model/R/non_linear_model_feb_2018/01b_plot_profile_make_fig_FY_wider/find_zero_crossing.py

\subsection{Numerical modelling results and discussion}\label{NumModResults}
% Appendix~\ref{NonLinearModel} describes our non-linear two-dimensional numerical model of the instability.

Summary: the modelled evolving flow
exhibits instability which is visually similar to that 
observed in the bucket, when initialized with a velocity profile similar to that
measured.
The emergent modelled preferred wavenumber (which is three in this instance) is the same as that observed, 
and the modelled growth rate is consistent with that observed.

% First talk about passive tracers...
A common visualization tool for water flows is to add coloured dye to some parts of the water. The dye acts as a passive ``tracer'': showing the movement without influencing the movement.
In our physical demonstration the tea leaves perform a similar function, except that they tend to congregate in the vortices due to bottom boundary effects which are not simulated in our numerical model.
It is relatively simple to add a simulation of a numerical passive tracer to the numerical model, and the result gives an intuitive sense of the behavior,
because the tracer looks like phenomena which can be seen in the real world --- for example the patterns seen in creamer added to coffee.

A video clip of the simulation, VideoS2,
available at 
 \href{https://vimeo.com/399593365}{\texttt{https://vimeo.com/399593365}} (password: Welcome123),   %  4-panel anim (.../28/...)
shows a slow motion (one-fifth speed) animation of some aspects of the simulation, 
concentrating on the most interesting period of flow evolution.
The top left-hand panel shows
our tracer. %during the growth part of the instability. 
The tracer starts out as a sinusoidal pattern of
4 radial cycles and 7 azimuthal cycles %,                           as seen in the first frame of the animation.
We chose 7 cycles to ensure that we did not prejudice the visual emergence of three vortices.
Green shows positive values of tracer and purple shows negative values of tracer.
The tracer diffuses, so we  reset it at about 4.5 seconds to capture the most interesting part of the evolution.

A alternative intuitive visualization is shown in the top right-hand panel: a cloud of dots moving with the flow. 
Physically these are 
analogous to paper dots floating on the surface of the water in the bucket. The initial positions of the dots are chosen randomly. 
A short ``tail'' shows the recent track of each dot. 
The passive tracer of the top left panel is shown again faintly in gray for comparison. 

Neither the tracer nor the dots are subject to the process which tends to concentrate the tea leaves 
in the vortices in the physical demonstration, and
the formation of three separate vortices is seen more clearly in an animation of the stream function, 
shown by color-filled contours in the bottom left panel.
You can see that three sets of closed streamlines appear, corresponding to three vortices. 
Thus the three areas of closed streamlines correspond to the three areas in which the tea leaves tend to congregate in the physical experiment.
% Also shown by the colored areas in the same panel is the radial component of velocity. Yellow shows outward movement and blue shows inward.
% Comparing this with the streamlines aids in interpretation of the streamlines. 
% For example you can see that the streamlines are near-radial where the radial velocity is large (inward or outward). 
%
The bottom right panel shows the dots with their ``tails'' again, this time overlain on contours of the stream function (streamlines).
By watching this panel closely (and, if possible, stopping the animation after the three sets of closed streamlines have appeared and stepping through one frame at a time)
you can see for yourself that the instantaneous movement of the particles is always along the streamlines, even
though the particle paths are not identical to streamlines (because the streamlines evolve).

Video clip VideoS3,
available at
\href{https://vimeo.com/279323035}{\texttt{https://vimeo.com/279323035}} (password: Welcome123),   %  long animation of stream function
shows the evolution of the stream function in real time over the entire 30-second-long simulation.
The sense of the imposed initial rotation was
clockwise in both the physical demonstration and the simulation.

\begin{figure*}[t]
% something about using a diff syntax for wider figures? Can't seem to find it now, but see example at /net/home/h03/hadto/bucket/paper/BAMS/two_column01/howardTwoCol.tex
% That may only apply to using the author's own use version, though? Perhaps this is OK for the actual submission? Seems to work fine for the single-column version.
% \noindent\includegraphics[width=27pc,angle=0]{fig-GI.png}\\
% ~hadto/bucket/model/R/non_linear_model_feb_2018/07_wider_profile/fig_GI.eps
  \noindent\includegraphics[width=40pc,angle=0]{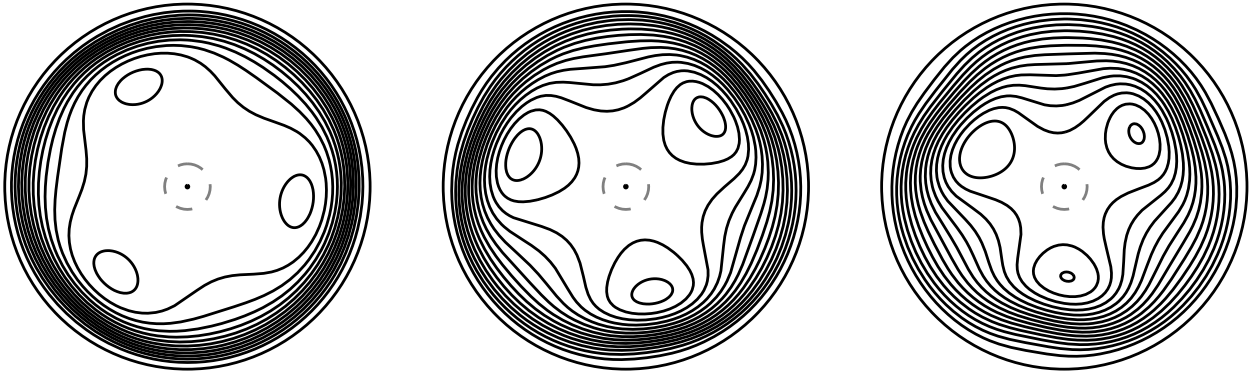}\\
  \caption{ Streamlines (contours of the stream function) of the numerical simulation at 6.75, 8.25 and 9.75 seconds (from left to right). Contour
interval 0.0005 \protect{m$^2$~s$^{-1}$}. The dashed line shows the boundary of the center blob (see \protect{section~\ref{blob}}).
The whole tripolar pattern rotates clockwise: the vortex at about ``11 o'clock'' in the left-hand panel is at about 
``10 o'clock'' in the center panel and about ``6 o'clock'' in the right-hand panel (see video clips VideoS2 [URL will be inserted by AIP] and VideoS3 [URL will be inserted by AIP]).}
  \label{figPsi1}
\end{figure*}

\begin{figure*}[t]
% \noindent\includegraphics[width=27pc,angle=0]{fig-GK-better03.png}\\
% /net/home/h03/hadto/bucket/model/R/non_linear_model_feb_2018/07_wider_profile/fig-GK-centred01.eps
  \noindent\includegraphics[width=40pc,angle=0]{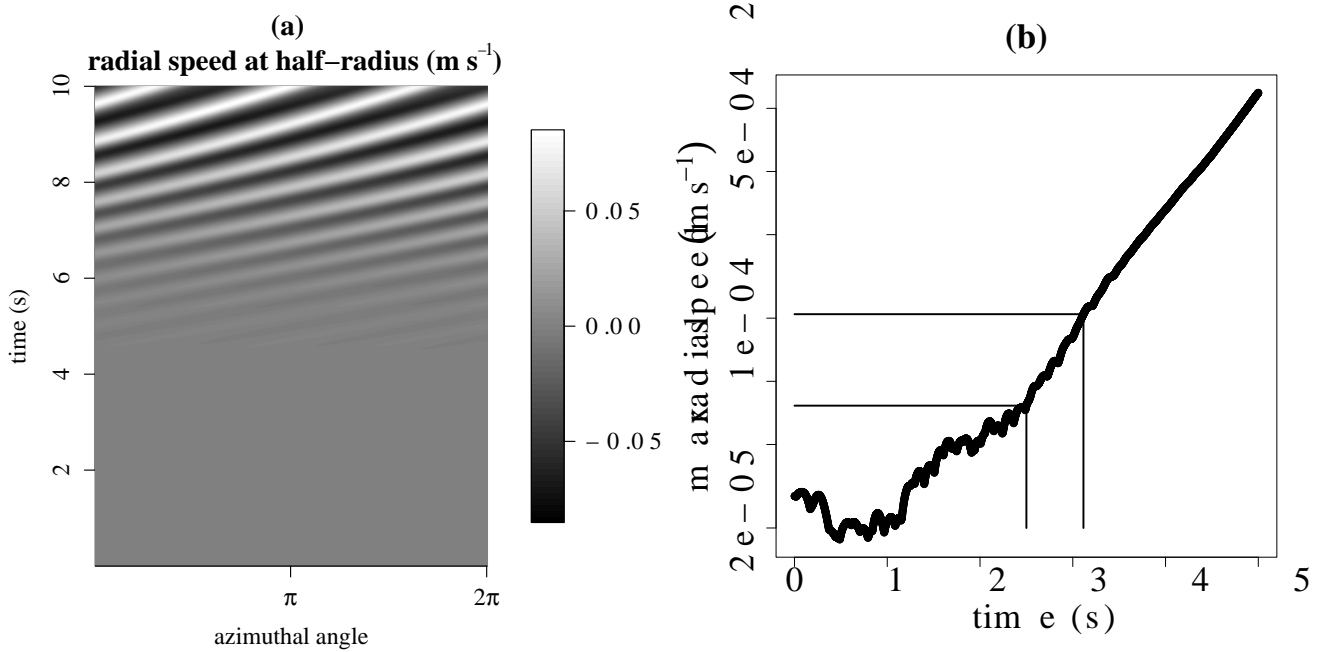}\\
  \caption{(a)  Hovmoller diagram illustrating growth of radial velocity at half-radius.
           (b) Time series of maximum radial speed at mid-radius, on a logarithmic scale. E-folding time of the most rapid growth is approximately 0.6 seconds.}
  \label{figHovmoller}
  \label{figMax}
\end{figure*}

Figure~(\ref{figPsi1})  shows snapshots of the evolving simulated
stream function after the instability has begun to grow. 
To investigate the growth of the instability
during a particular simulation we focus on the radial velocity at approximately the mid radius (about 45~mm).
A Hovmoller diagram is a common way of plotting meteorological data. The axes are typically longitude or latitude (x-axis) and time (y-axis) with the value of some field represented through color or shading.
Figure~(\ref{figHovmoller}(a)) is a Hovmoller diagram of the radial velocity at mid radius, which illustrates the inception, growth, and
advection (movement by the flow) of the vortices arising from the instability.

To estimate the growth rate of the modelled instability we use the maximum radial
velocity at the mid radius as a simple metric of the strength of the instability. The evolution of this
metric is shown in Fig.~(\ref{figMax}(b)). Choosing a small section of the steepest part of the curve (delineated by the
straight lines on the plot) facilitates estimation of the growth rate. We find an e-folding time of
approximately 0.6 seconds. 
Growth rates in the experiment and model can be visually compared by comparing the video clips; they appear similar.
We do not have a precise measurement of the growth rate in the experiment, but the three vortices are well-established 
by time t=8.2 seconds on the video clip, unidentifiable at 6.2 seconds, and arguably detectable at around 7.2 seconds.
These observations are not inconsistent with the e-folding time of 0.6 seconds obtained from the model.

% A necessary criterion for symmetric overturning instability is that the angular momentum 
% decreases outward, and this is satisfied for our velocity profile in the curved case, outside of $r \approx 0.073$~m (see Fig.~(\ref{figProfiles})).
The model is quite basic, perhaps even crude by contemporary standards. 
However, a two-dimensional linear stability analysis
yields very similar results in terms of wavenumber and growth rate of the instability (see appendix~\ref{LinearStabilityAnalysis}).
The cross-corroboration of the three experiments (laboratory, numerical and analytical)
gives confidence in the numerical model results, in spite of its simplicity, 
and supports our suggestion that the instability is fundamentally two-dimensional.
The strong similarity between the behavior seen in the two-dimensional model and the observed behavior
suggests that, although a centrifugal instability may be involved in the creation of the profile we measured,
a two-dimensional process with negligible variation in the vertical direction --- we suggest the barotropic (shear-flow) instability 
associated with the change in sign of the radial gradient of vorticity --- can account for the evolution of the flow 
forwards in time from our axisymmetric measured profile.

% %  consistent with the linear analysis.
% \begin{figure}[t]
% % \noindent\includegraphics[width=19pc,angle=0]{fig-GJ-better02-notitle.png}\\
%   \noindent\includegraphics[width=19pc,angle=0]{HowardFig06.eps}\\
%   \caption{Time series of maximum radial speed at mid-radius, on a logarithmic scale. E-folding time of the most rapid growth is approximately 0.6 seconds.}
%   \label{figMax}
% \end{figure}

\subsection{Modified numerical model: cyclic straight channel}\label{straightChannel}
The instability is also simulated when the curvature is artificially removed, modelling a hypothetical
straight channel with cyclic lateral boundaries.

A well-recognized advantage of numerical modelling is the ease of modifying the experiment. To show
that the existence of an instability depends on the velocity profile, and not on the curvature of the domain, we modified
the numerical simulation. The modified simulation consists of a straight two-dimensional domain with
cyclic boundaries in the primary ($x$) direction and walls bounding the other ($y$) direction. Thus the $y$
direction replaces the radial ($r$) direction of the previous simulation, and we choose the width (in the $y$
direction) of the domain to match the radius of the bucket. The velocity profile that was used to initialize
the previous simulation is here the initial $x$-direction velocity, and we choose the length (in the $x$
direction) of the domain to be approximately the circumference of the bucket (about 0.5 meters).
We use a rectilinear grid in place of the polar grid of the previous simulation, and remove any of
the terms describing effects associated with the curvature of the domain.
In order to study the effect of only the removal of the curvature, we retain the  
cyclic lateral boundary conditions.
This simulation exhibits a similar instability to that seen in the curved simulation (see Fig.~(\ref{figStraight})).
% (although a smaller wavenumber is selected by the flow in this instance). 

The cyclic lateral boundaries create a periodic domain and this quantizes the allowable wavenumbers.\citep{Vallis}
We also performed a simulation with cyclic lateral boundaries and a doubled length (in the $x$~direction). 
% I would explicitly say that the velocity profile/stream function remained unchanged and maybe remind wavelength=L/m.
The initial velocity profile remained unchanged.
The double-length simulation exhibited an instability with six nodes, compared to the three nodes of the standard length straight channel.
Since the wavelength is the channel length divided by the number of nodes, this finding (that the wavelength is the same in spite of doubling the channel length) is
consistent with the expectation that the 
selected azimuthal wavelength for the shear instability scales with the shear layer width,\citep{Vallis} rather than the available length.  % (Billant, pers comm, 2014).
In other words, the number of vortices seen in the bucket demonstration (typically three) depends on the width of the shear layer. 
By experimenting with different initial flicks and 
different periods of suspension before the bucket is set back down, different numbers of vortices may sometimes be produced. Three or four was our most frequent outcome.

Our straight-channel results suggest that the essential
requirement for the instability is the velocity profile: 
% the curvature of the domain modifies the response, but
a curved domain is not a requirement for the instability.

% \section{Upscale behavior}
% Given a sufficiently strong initial flick, if the flow is left to evolve, the multiple vortices ultimately merge
% into a single axisymmetric vortex centered on the middle of the bucket. This is reminiscent of the upscale
% cascade, which is a feature of two-dimensional flows. A similar merging is exhibited in the numerical model.
% However, we do not analyse this  here,
% beyond remarking on the visual similarity.

\section{Summary and Conclusions}
% Polygonal patterns were observed in a classroom as an accidental by-product of an Ekman pumping demonstration. 
% The linear analysis of an experimental velocity profile shows that these observations are in agreement with barotropic instability theory. 
The highlight of this paper is that it presents a very simple, very low-budget classroom demonstration of dynamic instability in a rotating fluid.
% The demonstration is reproducible in the classroom with only a bucket of water, a piece of string, and some used tealeaves. 
% Given its simplicity, this demonstration can be ported to outreach activities and 
% play a role in demystifying observations 
% of polygonal patterns in geophysical flows.
The demonstration can be used to introduce the concept of fluid-dynamic instability in a classroom context, or, given its simplicity,
it can be ported to outreach activities. Thanks to the low cost, multiple sets of the equipment can readily be provided to
create a participatory activity. 
The activity can play a role in demystifying observations 
of polygonal patterns in geophysical flows.
The breaking symmetry of an initially axially-symmetric flow in water is  visualized using tealeaves or coffee grounds.
Consistency of the observed phenomenon with two-dimensional toy models suggests that the instability is essentially two-dimensional 
in nature and furthermore the existence of the instability does not depend on the curvature of the domain, but 
on the shape of the initial velocity profile.

We close with a speculation.
Given a sufficiently strong initial flick, if the flow is left to evolve, the multiple vortices ultimately merge
into a single axisymmetric vortex centered on the middle of the bucket. This is reminiscent of the upscale
cascade, which is a feature of two-dimensional flows. 
Given the similarity between the instability in the early stages of our experiment and our numerical and analytical models, which are two-dimensional, could this 
merging be a manifestation of the upscale cascade?
Further work would be needed 
before a more robust assertion about this could be made, 
but we note that a similar merging is exhibited in the late stages of flow in our numerical model.
% However, we do not analyse this  here,
% beyond remarking on the visual similarity.

% We also note that subsequent behavior is visually reminiscent of the upscale cascade. However, further work would
% be needed to establish more than a visual similarity.

\section{Supporting Information}\label{Links}
%\section{Video clips of alternative configurations}\label{Links}

The text is supported by the following supplementary information:

%appendices.pdf: Two appendices and supplementary figure S2, all referenced in the main text.
%appendices.pdf: Supplementary figure S2 and an appendix, all referenced in the main text.

Video clip of demonstration:  VideoS1:
\href{https://vimeo.com/278481176}{\texttt{https://vimeo.com/278481176}} (password: Welcome123)  % physical demo

Video clip of numerical simulation (slow motion): VideoS2: 
\href{https://vimeo.com/399593365}{\texttt{https://vimeo.com/399593365}} (password: Welcome123),   %  4-panel anim (.../28/...)

Video clip of numerical simulation (real time): VideoS3:
\href{https://vimeo.com/279323035}{\texttt{https://vimeo.com/279323035}} (password: Welcome123)   %  long animation of simulation
% Video clip of numerical simulation (passive tracer; slow motion): \href{https://vimeo.com/384555742}{\texttt{https://vimeo.com/384555742}} (password: Welcome123)   %  slow motion animation of tracer: key part

Video clip of taller, narrower configuration (120 mm diameter by 100 mm deep): VideoS4:
\href{https://vimeo.com/323816117}{\texttt{https://vimeo.com/323816117}}  (password: Welcome123)

Video clip of configuration with rigid lid: VideoS5:
\href{https://vimeo.com/323821255}{\texttt{https://vimeo.com/323821255}}  (password: Welcome123)

\appendix
% May need to update this if figure numbers change !!
\section{Stream function of the simulated straight cyclic channel}\label{psiStraight}
\begin{figure}[H]
% \noindent\includegraphics[width=27pc,angle=0]{fig-GI-straight-long.png}\\
% \noindent\includegraphics[width=27pc,angle=0]{fig_GI_bucket_as_straight_channel_rectangular_long_for_comparison.png}\\
% see ~hadto/bucket/model/R/non_linear_model_feb_2018/11_as_08_but_length_based_on_outer_circumf/README
  \noindent\includegraphics[width=27pc,angle=0]{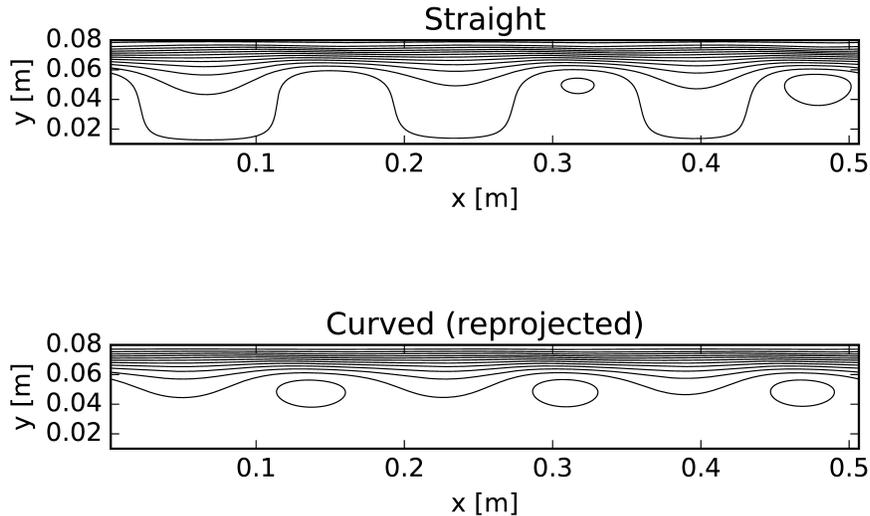}\\
% May need to update this if figure numbers change !!
  \caption{Top: Stream function of the simulated straight cyclic channel flow at 6.75 seconds.  Contour interval
0.0005~\protect{m$^2$ s$^{-1}$}.
Bottom: as top, but data from the curved simulation reprojected onto a straight plot for ease of comparison.
For the given velocity profile,
these two plots illustrate the result that although curvature modifies the behavior
(the two plots are not identical), it is not a requirement
for instability
(both configurations exhibit instability). }
% simulation). Bottom: contours of the stream function reprojected onto a polar plot for ease of
% comparison with \protect{Fig.~(\ref{figPsi1}).} }
  \label{figStraight}
\end{figure}

\section{Linear stability analysis}\label{LinearStabilityAnalysis}    %% Appendix B   This is the linear analysis description and results.
% ?? $v \upsilon$
Our stability analysis follows quite closely that described by Barbosa Aguiar et al.\citep{Aguiar}
In our analysis, we
consider only the barotropic mode; we ignore friction; and we assume
constant depth, implying a beta parameter of zero.
Thus we have a two-dimensional problem with
vorticity conserved following the flow.
We write all the dependent variables as the sum of a fixed base state (indicated by an overline) and a small perturbation (indicated by a prime), e.g. $u=\overline{u} + u'$.
We linearize about the base state  with
base-state azimuthal speed $\overline{u}=\overline{u}(r)$ (we use the fitted analytical profile shown in Fig. 4 of the main text) to give
\begin{equation}
\frac{\partial \zeta'}{\partial t} + \frac{\overline{u}}{r}\frac{\partial \zeta'}{\partial \theta} + v \frac{\partial \overline{\zeta}}{\partial r} = 0,  
\label{vort_eq}
\end{equation}
where  $r $ is radius, $\theta $ is
the azimuthal angle (in radians), $t $ is time,
$\overline{\zeta}$ is the base-state vorticity, and $\zeta'$ is the perturbation vorticity.
$v$ is the radial component of velocity, which is of perturbation order, but we can omit the prime because $\overline{v}=0$.
It is convenient to define $\theta $ as
increasing in the direction of positive $u $. We introduce a stream function $\psi $ such that $u = - {\partial \psi}/{\partial r}$,
$v = {\partial \psi}/{r\partial \theta}$ 
and 
$\zeta = \nabla^2 \psi$.
% # $v$ is the radial velocity, which is a perturbation order quantity.

% The solution to Eq.~\ref{vort_eq} is the sum of Fourier modes of the form
The solution to Eq.~(\ref{vort_eq}) can be expressed in terms of the sum of Fourier modes of the form
\begin{equation}
\psi ' = \mathrm{Re}\  \tilde{\psi}(r, k) \exp{\{i(\sigma t - k \theta)\}},
\end{equation}
where $\tilde{\psi}(r, k)$ is independent of $\theta$, $i\sigma$ is a (possibly complex) growth rate and $k$ is wavenumber $(k=1,2,3...)$.
Since the problem is linear, the modes do not interact and we can consider them separately.

We seek to identify the fastest-growing modes. We discretize in the $r$ direction and eventually arrive at
\begin{equation}
\sigma \pmb{\mathsf{D}} \phi - k\left[ \frac{\overline{u}}{r}\right] \pmb{\mathsf{D}} \phi - k \left[ \frac{d \overline{\zeta}}{r dr} \right] \phi = 0
\label{eqMat1}
\end{equation}

where $\pmb{\mathsf{D}}$ is a matrix representing the discrete form of the operator

\begin{displaymath}
\frac{d^2}{dr^2} - \frac{k^2}{r^2} + \frac{1}{r} \frac{d}{dr}  
\end{displaymath}

and $\phi$ is a vector representing the discrete form of $\tilde{\psi}(r, k)$. Quantities in square brackets are diagonal
matrices of known discrete values. Eq.~(\ref{eqMat1}) can be further manipulated to give a set of eigenvalue
problems

\begin{equation}
\frac{\sigma}{k}\phi = \pmb{\mathsf{M}}\phi, \hspace{1cm} k=1,2,3...
\end{equation}

where

\begin{displaymath}
\pmb{\mathsf{M}} = \pmb{\mathsf{M}}(k) = \pmb{\mathsf{D}}^{-1}\left( \left[ \frac{\overline{u}}{r}\right] \pmb{\mathsf{D}} + \left[ \frac{d \overline{\zeta}}{r dr} \right] \right)
\end{displaymath}

\begin{figure}[t]
% \noindent\includegraphics[width=19pc,angle=0]{fig-GP-10.png}\\
  \noindent\includegraphics[width=19pc,angle=0]{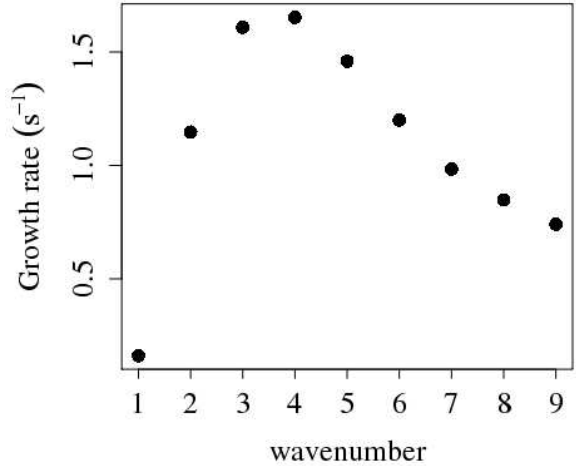}\\
  \caption{Growth rate, \protect{$i\sigma$} vs wavenumber, \protect{$k$}, for the profile shown by the black line in Fig. (4) of the main text}
  \label{figGrowth}
\end{figure}

A careful treatment of the boundary conditions is important in the non-linear model (see main text)
but we have found that the results of our linear model are not sensitive to the use of Dirichlet
or Neumann boundary conditions. The growth rate of the fastest-growing radial eigenfunction for each
%avenumber up to nine is shown in Fig.~(\ref{figGrowth}). The e-folding time (which is the reciprocal of the growth
wavenumber up to nine is shown in 
%
% May need to update this if figure numbers change !!
Fig.~(\ref{figGrowth}).
The e-folding time (which is the reciprocal of the growth
rate) for wavenumbers 3 or 4 is about 0.6 seconds (as in the numerical model: see 
% Need to update this if section numbers change !!
% This should refer to the section called: Numerical modelling results and discussion
section~\ref{NumModResults}
% section IV.B  
%
of the main text).

\begin{acknowledgments}
This work was supported by the Met Office Hadley Centre Climate Programme funded by BEIS and Defra. 

Thanks to Grae Worster and an anonymous referee of an earlier draft of this paper for pointing out the centrifugal instability of the boundary layer mentioned in section~\ref{nature}.
We acknowledge the contribution of the anonymous referees who helped to improve earlier drafts of this article. 
Thanks to Simon Hammett for the stop-motion filming, which facilitated an estimation of the velocity profile.
Thanks to Mike Cullen, Paul Billant, Michael McIntyre,  Mike Bell, Richard Wood, Adam Scaife, Eddy
Carrol, Andy White, Nigel Wood, David Thomson, Philip Brohan and Matt Palmer for various helpful discussions, emails, advice and encouragement. 

\copyright British Crown Copyright 2020, Met Office

This article may be downloaded for personal use only. Any other use requires prior permission of the author and AIP Publishing. This article appeared in American Journal of Physics 88, Issue 12, 1041 (2020) and may be found at \href{https://doi.org/10.1119/10.0002438}{\texttt{https://doi.org/10.1119/10.0002438}}

\end{acknowledgments}

\bibliographystyle{ajpdjg}
\bibliography{howard_arxiv.bib}
%%
%% URLs and DOIs can be entered in your BibTeX file as:
%%
%% URL = {http://www.xyz.org/~jones/idx_g.htm}
%% DOI = {10.5194/xyz}

\end{document}